\documentclass[fleqn,usenatbib]{mnras}

\usepackage[T1]{fontenc}

\DeclareRobustCommand{\VAN}[3]{#2}
\let\VANthebibliography\thebibliography
\def\thebibliography{\DeclareRobustCommand{\VAN}[3]{##3}\VANthebibliography}

\usepackage{graphicx}	%
\usepackage{amsmath}	%
\usepackage{amssymb}	%
\usepackage{newtxtext,newtxmath}

\usepackage{booktabs}
\usepackage{siunitx} %
\DeclareSIUnit[]\solarmass
{\text{\ensuremath{\textup{M}_{\odot}}}}
\DeclareSIUnit[]\solarluminosity
{\text{\ensuremath{\textup{L}_{\odot}}}}
\DeclareSIUnit[]\solarradius
{\text{\ensuremath{\textup{R}_{\odot}}}}
\DeclareSIUnit[]\year
{\text{yr}}
\DeclareSIUnit[]\au
{\text{AU}}
\DeclareSIUnit[]\parsec
{\text{pc}}
\DeclareSIUnit[]\erg
{\text{erg}}
\DeclareSIUnit[]\arcsecond
{\text{as}}

\newcommand{\ts}{\textsuperscript}

\newcommand{\swr}{\ensuremath{_{\text{WR}}}}
\newcommand{\sob}{\ensuremath{_{\text{OB}}}}
\newcommand{\rms}[1]{\ensuremath{_{\text{#1}}}}
\newcommand{\mdot}{\dot{\text{M}}}
\newcommand{\dsep}{d\rms{sep}}

\title[Dust growth simulations of WR140]{Exploring dust growth in the episodic WCd system WR140}

\author[J. W. Eatson, J. M. Pittard \& S. Van Loo]{
J. W. Eatson,$^{1,2}$\thanks{E-mail: \texttt{\href{mailto:py13je@leeds.ac.uk}{py13je@leeds.ac.uk}}}
J. M. Pittard$^{1}$
and
S. Van Loo$^{1,3}$
\\
$^{1}$School of Physics and Astronomy, University of
       Leeds, Woodhouse Lane, Leeds LS2 9JT, UK\\
$^{2}$Department of Physics and Astronomy, The University of Sheffield, Hicks Building, Hounsfield Road, Sheffield, S3 7RH, UK\\
$^{3}$Department of Applied Physics, Ghent University, Sint-Pietersnieuwstraat 41, Technicum blok 4
9000 Gent, Belgium
}

\date{Accepted XXX. Received YYY; in original form ZZZ}

\pubyear{2022}

\begin{document}
\label{firstpage}
\pagerange{\pageref{firstpage}--\pageref{lastpage}}
\maketitle

\begin{abstract}
  \noindent
  The wind collision region (WCR) in a colliding wind binary (CWB) is a particularly violent place, as such, it is surprising that it is also a region where significant quantities of interstellar dust can form.
  In extreme cases, approximately 30\% of the total mass loss rate of a system can be converted into dust.
  These regions are poorly understood, as observation and simulation of these systems are difficult.
  In our previous paper we simulated dust growth in CWB systems using an advected scalar model and found our model to be suitable for qualitative study.
  For this paper we simulated the periodic dust forming CWB (WCd) system WR140 with our dust model, to determine how dust growth changes over the systems periastron passage.
  We found that dust production increases significantly at periastron passage, which is consistent with IR emission of the surrounding dusty shell.
  We also find that the dust production rate of the system decreases rapidly as the stars recede from each other, though the rate of decrease is significantly lower than the rate of increase during periastron passage.
  This was found to be due to strong cooling and its associated thermal instabilities, resulting in cool, high-density pockets of gas in the WCR where dust forms.
  The WCR also shows a degree of hysteresis, resulting in a radiative post-shock flow even when the stars are separated enough for the region to behave adiabatically.
\end{abstract}


\section{Introduction}

Dust formation in massive star binary systems is a particularly fascinating subject.
Considering the immense photon fluxes and strong shocks involved, these systems should not form dust at all, and yet in some systems all evidence points to the contrary.
Colliding wind binary (CWB) systems were first hypothesised to explain highly luminous and variable x-ray emission in systems such as V444 Cyg and $\gamma^2$ Vel \citep{prilutskii_x_1976}.
These extremely bright emissions were found to be due to stellar wind collision with shock velocities of the order $10^3 \, \si{\kilo\metre\per\second}$.
The variability in x-ray emission can be explained if the phenomena occurs due to the orbit of a binary system, such as the Wind Collision Region (WCR) being occluded by the outflowing stellar winds and being occluded by the stars themselves.
The system can also have an eccentric orbit, changing the shock strength as the orbital separation, $d\rms{sep}$, varies.
Despite this dust-hostile environment, CWB systems containing a Wolf-Rayet carbon phase star (WC) have been observed producing copious quantities of dust (so-called WCd systems; see \cite{Williams1995}).
These systems typically convert around $1\%$ of the stellar wind into dust a short time after wind collision; in more prolific systems such as WR104 up to $36\%$ of the Wolf-Rayet (WR) outflow is converted into dust \citep{lauRevisitingImpactDust2020}.
This corresponds to dust production rates on the order of $10^{-6} \, \si{\solarmass\per\year}$, rivalling other profuse dust producing phenomena such as AGB stars.

WCd systems can be sub-categorised further, into persistent, variable and episodic dust forming systems.
Persistent systems, such as WR104 \citep{tuthill_dusty_1999}, produce dust at a constant rate, and as such produce extreme quantities of dust, as well as well-defined pinwheel patterns if the system is viewed face-on.
Episodic systems, meanwhile, only produce dust for a limited period before entering a period of dormancy; this pattern is cyclical, and is predictably periodic.
A good example of such an episodic system is WR140 \citep{williamsMultifrequencyVariationsWolfrayet1990}, the subject of this paper.
Variable systems have some characteristics of these two sub-types, having a distinct variability without a period of dust producing dormancy, such as WR98a \citep{monnierPinwheelNebulaWR1999}.

Whether a system is persistent, variable or episodic is based on the system's orbital eccentricity.
Highly eccentric systems appear to form episodic systems, with the ``active'' dust production period occurring immediately after periastron passage, and a relatively short time thereafter.
Meanwhile, persistent and variable systems have been observed to have more circular orbits, suggesting that the effect of a change in system separation distance, $d\rms{sep}$, has a role in dust formation.
The initial mechanism behind dust formation is not well understood.
Whilst nascent amorphous carbon dust grain cores can condense within the photosphere of WC7-9 stars,
these grain cores would be vaporised by the UV flux of both stars.
However, within the WCR these grains appear to flourish.
Observations of these systems show that infrared excess in wavelengths associated with amorphous grains is detected almost exclusively within the post-shock WCR \citep{soulainSPHEREViewWolfRayet2018}.
Observations also indicate that dust formation occurs rapidly and close to the system, this requires strong radiative cooling for the immediate-post shock temperature to reduce from $\sim 10^7 \, \si{\kelvin}$ to $\sim 10^4 \, \si{\kelvin}$
\citep{williamsInfraredPhotometryLatetype1987,williamsMultifrequencyVariationsWolfrayet1990}.

As such, dust formation appears to be encouraged in the WCR through a multitude of factors:

\begin{itemize}
  \item Strong radiative cooling produces clumps of cool, high density gas where dust can rapidly grow.
  \item The high density of the post-shock WCR results in a high collision rate between carbon atoms and dust grains.
  \item The high density also shields nascent dust grains from the bulk of the UV emission from the stars.
  \item The rapid cooling in the immediate post-shock environment reduces gas-grain sputtering.
\end{itemize}

\noindent
The dust formation can also be influenced by orbital separation, velocity shear and momentum ratio imbalance between the winds, producing variability on the timescale of a single orbit, or $t\rms{dyn} \ll P$.

\begin{table}
  \centering
  \resizebox{\linewidth}{!}{
  \begin{tabular}{lllllll}
    \hline
    & \multicolumn{2}{c}{Persistent} & \multicolumn{2}{c}{Variable} & \multicolumn{2}{c}{Episodic} \\ \cline{2-7} 
    & Total & Example & Total & Example & Total & Example \\ \hline
   WC4 & 1 & WR19 & 0 & --- & 0 & --- \\
   WC5 & 0 & --- & 0 & --- & 1 & WR47C \\
   WC6 & 1 & WR124-10 & 0 & --- & 0 & --- \\
   WC7 & 3 & WR102-22 & 0 & --- & 4 & WR140 \\
   WC8 & 6 & WR13 & 1 & WR48a & 3 & WR122-14 \\
   WC9 & 45 & WR104 & 6 & WR98a & 1 & WR75-11 \\ \hline
   Total & 56 &  & 7 &  & 9 &  \\ \hline
  \end{tabular}
  }
  \caption[Confirmed WCd systems]{WCd systems with a known spectral type and dust formation type from the Galactic Wolf Rayet Catalogue \citep{rossloweSpatialDistributionGalactic2015}. Systems with uncertain spectral types are not included, while systems labelled ``d'' are included within the ``persistent'' category for their associated spectral type.}
  \label{tab:p2-wc-summated-list}
\end{table}

WCd systems are comparatively rare.
Out of 106 confirmed WR binaries, only 9 are categorised as episodic WCd systems
(Table \ref{tab:p2-wc-summated-list}).
As these systems have a typical distance on the order of $1-10 \, \si{\kilo\parsec}$, observation of WCds is difficult.
Whilst these systems can be observed and the dusty WCR can be resolved, observation of the innermost, immediate post-shock dust forming region is not possible at this distance.
As such, numerical simulation is necessary to determine dust formation in WCd systems.
A contemporary example of such simulations is \cite{hendrix_pinwheels_2016}, though the evolution of dust grains through cooling, growth and sputtering was not performed.
In this paper we present a numerical simulation of the archetypical episodic WCd system WR140 with a co-moving dust model simulating grain growth and sputtering through gas-grain collisions.
This simulation covers a temporal slice of the orbit of WR140 from phase $\Phi = 0.95$ to $\Phi = 1.10$, or the period immediately prior to and after periastron passage.
We will discuss our methodology in Section \ref{sec:paper-2-methodology}, with a particular emphasis on our dust model in Subsection \ref{sec:dust-model}.
Afterwards we will discuss the simulation and WR140 system parameters, as well as our data collection techniques in Section \ref{sec:paper2-wr140}.
Finally, we will discuss our results and conclude in Sections \ref{sec:p2-results} and \ref{sec:p2-conclusion}.

\section{Methodology}
\label{sec:paper-2-methodology}

The periodic dust forming system WR140 was simulated using a fork of the Athena++ hydrodynamical code \citep{stoneAthenaAdaptiveMesh2020}.
A series of modifications were implemented to simulate binary system orbits, stellar wind outflows and dust evolution.
These simulations were conducted in 3D in a Cartesian co-ordinate system.
The code solves a Riemann problem at each cell interface to determine the time-averaged values at the zone interfaces, and then solves the equations of hydrodynamics:

\begin{subequations}
  \begin{align}
    \frac{\partial\rho}{\partial t} & +\nabla \cdot \left(\rho \boldsymbol{u}\right) = 0 , \\
    \frac{\partial \rho \boldsymbol{u}}{\partial t} & + \nabla \cdot \left(\rho \boldsymbol{u} u + P \right) = 0, \\
    \frac{\partial \rho \varepsilon}{\partial t} & + \nabla \cdot \left[ \boldsymbol{u} \left( \rho\varepsilon + P \right) \right] = \mathcal{L}\rms{T} , 
  \end{align}
\end{subequations}

\noindent
where $\varepsilon$ is the total specific energy ($\varepsilon = \boldsymbol{u}^2/2 + e/\rho $), $\rho$ is the gas density, $e$ is the internal energy density, $P$ is the gas pressure and $u$ is the gas velocity.
In order to simulate radiative losses, the parameter $\mathcal{L}\rms{T}$ is included, which is the energy loss rate per unit volume from the fluid due to gas and dust cooling.

Spatial reconstruction using a piecewise linear method was performed, while the simulation was run with one of two numerical integrators, depending on the system stability.
The integrators used were a 3\ts{rd} order accurate Runge-Kutta integrator (\texttt{rk3}), as well as a 4\ts{th} order accurate, 5-stage, 3 storage register strong stability preserving Runge-Kutta integrator (\texttt{ssprk5\_4}; \citet{ruuthHighOrderStrongStabilityPreservingRungeKutta2005}).
The \texttt{ssprk5\_4} integrator was found to be approximately 60\% slower, but significantly more stable.

Several passive scalars are utilised to model wind mixing and dust evolution, which are transported by the fluid.
For a given scalar, $i$, the scalar is advected through the following equation:

\begin{equation}
  \frac{\partial}{\partial t} \left( \rho C_i \right) + \nabla \cdot \left( C_i \rho \mathbf{u} \right) = 0 ,
  \label{eq:p2-scal}
\end{equation}

\noindent
where $C_i$ is the scalar quantity.
The scalar quantities tracked in this model are:

\begin{itemize}
  \item $C$, the wind ``colour'', or mixing fraction (see Eq. \ref{eq:p2-c}).
  \item $z$, the dust-to-gas mass fraction (see Eq. \ref{eq:p2-z}).
  \item $a$, the grain radius in microns.
\end{itemize}

\noindent
As there is no diffusion coefficient in Eq. \ref{eq:p2-scal}, these quantities can be interpreted as explicitly co-moving with the fluid.

Stellar winds are simulated by modifying the density, $\rho_R$, momentum, $p_R$, and energy, $E_R$ in a small region around both stars.
Winds flow from this ``remap'' region at the stars wind terminal velocity, $v^\infty$. Remap zone parameters are calculated with the formulae

\begin{subequations}
  \begin{align}
    \rho_R & = \frac{\mdot}{4 \pi r^2 v_R} , \\
    p_R    & = \rho_R v_{R} , \\
    E_R    & = \frac{P_R}{\gamma - 1} + \frac{1}{2} p_R v_R ,
  \end{align}
\end{subequations}

\noindent
where $P_R$ is the cell pressure ($P_R = \rho_R k_\text{B} T_w / \mu m_\text{H}$), $T_w$ is the wind temperature, $\mu$ is the mean molecular mass, $m_\text{H}$ is the mass of a hydrogen atom, $v_R$ is the wind velocity as it flows radially from the center of the ``remap zone'' and $r$ is the distance from the current cell to the centre of the remap zone.
This method produces radially out-flowing winds from the star with an expected density and velocity.

Line driving and wind acceleration effects are not simulated;
instead, winds are instantaneously accelerated to their terminal velocity.
Gravitational forces are not calculated as the winds are assumed to be travelling in excess of the system escape velocity.
In the case of systems with a very small orbital separation this could result in collision with a higher than anticipated wind velocity.
Therefore, calculating the line driving force would be a useful addition in the future when simulating close-orbit persistent dust forming systems such as WR104.

Athena++ utilises Message Passing Interface (MPI) parallelism.
The numerical problem is broken into blocks, which are distributed between processing nodes on a High Performance Compute (HPC) cluster.
The block size is variable, but for this simulation a block size of $40\times 40 \times 10$ cells in $XYZ$ was found to be optimal.
Static mesh refinement is used to increase the effective resolution of the simulation.
A region around the orbital path of the stars in the simulation is refined to a higher resolution, with gradual de-refinement as a function of distance from this refinement region.
This refinement is covered in more detail in Section \ref{sec:p2-sysparameters}.

\subsection{Radiative cooling}

Cooling is simulated via the removal of thermal energy from a cell at each time-step.
A cooling rate for radiative emission from the stellar winds, $\mathcal{L}\rms{g}$, is calculated and integrated using a sub-stepping Euler method.
The number of sub-steps is determined by the estimated cooling timescale of the cell.
Cooling is calculated via individual lookup tables from each wind.
These lookup tables contain the normalised emissivity, $\Lambda\rms{w}(T)$ at a logarithmically spaced series of temperatures from $10^4 \, \si{\kelvin}$ to $10^9 \, \si{\kelvin}$.
The cooling rate is determined for a cell by calculating the cell temperature and estimating $\Lambda\rms{w}(T)$ using linear interpolation between the nearest emissivity values in the lookup table.
As the two winds have significantly different abundances and can be thoroughly mixed in the WCR, we calculate an emissivity value for gas in a particular cell with the equation

\begin{equation}
  \Lambda\rms{g}(T) = C\Lambda\rms{w,WR}(T) + (1-C)\Lambda\rms{w,OB}(T) , 
  \label{eq:p2-c}
\end{equation}

\noindent
where $C$ is the wind ``colour'', or mixing fraction, where 1 is a pure WR wind and 0 is a pure OB wind.
$\mathcal{L}\rms{g}$, is then calculated through the equation

\begin{equation}
  \mathcal{L}\rms{g} = \left(\frac{\rho\rms{g}}{m\rms{H}}\right)^2 \Lambda\rms{g}(T),
\end{equation}

\noindent
where $\rho\rms{g}$ is the gas density and $m\rms{H}$ is the mass of hydrogen.
The lookup table was generated by mixing a series of cooling curves from a MEKAL calculation of elemental gasses \citep{mewe1995update}.
These curves were combined based on the elemental abundances in the WC \citep{sanderGalacticWCStars2012} and OB \citep{andersAbundancesElementsMeteoritic1989} winds.
To save calculation time, temperatures between $\SI{1e4}{\kelvin} < T \leq \SI{1.1e4}{\kelvin}$ are set to \SI{1e4}{\kelvin} as they are assumed to be either rapidly cooling or a part of the stellar wind outside of the WCR.
A minimum temperature of $10^4 \, \si{\kelvin}$ is defined by the simulation, is it is assumed that a radiating post-shock wind will tend to the temperature of the pre-shock wind, $T\rms{final} \rightarrow T\rms{pre-shock}$.

\subsection{Dust model}
\label{sec:dust-model}

In order to simulate dust evolution in WR140 a passive scalar dust model that simulates dust growth and destruction is included in the simulation.
The dust model operates on passive scalars, and as such simulates dust that is co-moving with the stellar wind.
Two scalars are used to describe dust in a cell, $a$, the grain radius in microns, and $z$, the grain dust-to-gas mass ratio

\begin{equation}
  z = \frac{\rho\rms{d}}{\rho\rms{g}},
  \label{eq:p2-z}
\end{equation}

\noindent
where $\rho\rms{d}$ is the dust density in the cell.
A number of assumptions are made in this dust model; for instance, the dust grains in the model are spherical, with a uniform density.
Dust grains are assumed to have a single size for each grid cell.
The number density of grains varies in the simulation volume, however the ratio of the grain number density is fixed and constant throughout the simulation.
As such, this model does not simulate grain fracturing.
Additional mechanisms for dust formation and destruction could also be implemented such as grain-grain agglomeration and photoevaporation.
A multi-fluid model with drag force coupling could also be implemented, however this is beyond the scope of this paper.

Dust grows through grain accretion using formulae described by \cite{spitzer_jr._physical_2008} where dust grains grow via low-velocity collisions with surrounding carbon atoms, causing them to accrete onto the surface of the dust grain.
Carbon is removed from the gas, reducing the cell gas density, while the corresponding dust density increases.
This ensures that mass is preserved in the simulation.
Assuming a single average grain size the rate of change in the grain radius in a cell due to accretion, $da\rms{acc}/dt$, is given by the equation:

\begin{equation}
  \frac{da\rms{acc}}{dt} = \frac{\xi \rho\rms{C} w\rms{C}}{4\rho\rms{gr}},
\end{equation}

\noindent
where $\xi$ is the grain sticking factor, $\rho\rms{C}$ is the carbon density ($\rho\rms{C} = X\rms{C} \rho\rms{g}$), $w\rms{C}$ is the Maxwell-Boltzmann RMS velocity for carbon ($w\rms{C} = \sqrt{3k\rms{B} T / 12m\rms{H}}$), $k\rms{B}$ is the Boltzmann constant and $\rho\rms{gr}$ is the grain bulk density.
The rate of change in the mass of a single grain due to accretion, $dm\rms{gr,acc}/dt$, is calculated with the formula:

\begin{equation}
  \frac{d m\rms{gr,acc}}{dt} = 4 \pi \rho\rms{gr} a^2 \frac{da\rms{acc}}{dt} = \pi \xi \rho\rms{C} w\rms{C} a^2, \\
\end{equation}

\noindent
A bulk density approximating that of amorphous carbon grains ($\rho\rms{gr} = \SI{3.0}{\gram\per\centi\metre\cubed}$) is used for this simulation.

Dust destruction through gas-grain sputtering is calculated using the \cite{drainePhysicsDustGrains1979} prescription.
Within a flow of gas number density $n\rms{g}$ a dust grain of radius $a$ has a grain lifespan, $\tau\rms{gr}$ of:

\begin{equation}
  \tau\rms{gr} = \frac{a}{da/dt} \approx \SI{3e6}{\year} \cdot \frac{a (\si{\micro\metre})}{n\rms{g}} \equiv \num{9.467e17} \cdot \frac{a}{n\rms{g}} .
\end{equation}

\noindent
This value is based on an average lifetime of carbon grains in an interstellar shock with a temperature of $\SI{1e6}{\kelvin} \leq T \leq \SI{3e8}{\kelvin}$ \citep{tielens_physics_1994,dwekCoolingSputteringInfrared1996}.
From this we find a rate of change in the grain radius due to sputtering of

\begin{equation}
  \frac{da\rms{sp}}{dt} = - \frac{a}{\tau\rms{gr}} = - \num{1.056e-18} \cdot n\rms{g}. 
\end{equation}

\noindent
The rate of change in the mass of a single dust grain due to sputtering, $dm\rms{gr,sp}/dt$, can then be calculated with a similar formula to the rate of change in grain mass due to accretion:

\begin{equation}
  \frac{d m\rms{gr,sp}}{dt} = 4 \pi \rho\rms{gr} a^2 \frac{da}{dt} = \num{-1.33e-17} \cdot n\rms{g} \rho\rms{gr} a^2 ,
\end{equation}

\noindent
The change in dust density is then calculated through the equation

\begin{equation}
  \frac{d \rho\rms{d}}{dt} = 
  \begin{cases}
    n\rms{d} \left(dm\rms{gr,acc} / dt \right) & \text{if } T \leq \num{1.4e4} \, \si{K} \\
    n\rms{d} \left(dm\rms{gr,sp} / dt \right)  & \text{if } T \geq \num{1.0e6} \, \si{K} , 
  \end{cases}
\end{equation}

\noindent
where $n\rms{d}$ is the dust grain number density.

Cooling via emission of photons from dust grains is also included in this model.
The rate of cooling is calculated using the uncharged grain case of the prescription described in \cite{dwek_infrared_1981}.
Grains are collisionally excited by collisions with ions and electrons, causing them to radiate.
Similarly to the gas/plasma emission model used, the emitted photons are not re-adsorbed by the WCR medium, causing energy to be removed from the simulation.
This therefore makes the assumption that the WCR is optically thin to far-infrared photons, which is observationally correct \citep{monnierKeckAperturemaskingExperiment2007,soulainSPHEREViewWolfRayet2018,callinghamAnisotropicWindsWolf2019}.
The grain heating rate, $H\rms{coll}$, in \si{\erg\per\second} for a dust grain is calculated with the formula:

\begin{equation}
  \label{eq:p2-grainheat}
  H = 1.26 \times 10^{-19} \frac{n\rms{g}}{A^{1/2}} a^2(\si{\micro\metre}) T^{3/2} h(a,T) , 
\end{equation}

\noindent
where $n\rms{g}$ is the gas number density,
$A$ is the mass of the incident particle in AMU,
$a(\si{\micro\metre})$ is the grain radius in microns,
$T$ is the temperature of the ambient gas,
and $h(a,T)$ is the effective grain heating factor.
Individual heating rates for hydrogen, helium, carbon, nitrogen and oxygen are calculated, in order to calculate the total ion collisional heating, $H\rms{coll}$:

\begin{equation}
  H\rms{coll} = H\rms H + H \rms{He} + H\rms C + H\rms N + H\rms O .
\end{equation}

\noindent
The effective grain heating factor for each element is calculated via the equation:

\begin{equation}
  h(a,T) = 1 - \left( 1 + \frac{E^*}{2 k\rms{B} T} \right) e^{- E^* / k\rms{B} T} ,
\end{equation}

\noindent
where $E^*$ is the critical energy required for the particle to penetrate the dust grain (Table \ref{tab:p2-criticalenergy}).
The rate of heating due to electron-grain collisions, $H\rms{el}$, is similar to Eq. \ref{eq:p2-grainheat}.
The grain heating factor for electron collisions, $h\rms{e}$, is calculated via an approximation rather than the exact calculation in the case of baryonic matter.
$h\rms{e}$ is estimated through the following conditions:

\begin{equation}
  \begin{alignedat}{3}
    h\rms{e}(x^*) & = 1 ,                && ~~ x^* > 4.5, \\
             & = 0.37{x^*}^{0.62} , && ~~ x^* > 1.5 , \\
             & = 0.27{x^*}^{1.50} , && ~~ \text{otherwise,}
  \end{alignedat}
\end{equation}

\noindent
where $x^* = \num{2.71e8} a^{2/3} (\si{\micro\metre})/T$.
This approximation differs from the integration method by less than 8\% while being 3 orders of magnitude faster.
Excitation due to grain-grain collisions were not modelled, due to the limitations of the passive scalar model.
In order to calculate the change in energy due to dust cooling, we find the radiative emissivity for dust, $\Lambda\rms{d}(T,a)$, to be

\begin{equation}
  \mathcal{L}\rms{d} = n_\text{d} (H\rms{coll} + H\rms{el}) ,
\end{equation}

\noindent
and added to the gas/plasma energy loss rate, such that the total energy loss rate is:

\begin{equation}
  \mathcal{L}\rms{T} = \mathcal{L}\rms{g} + \mathcal{L}\rms{d} .
\end{equation}

\begin{table}
  \centering
  \begin{tabular}{ll}
    \hline
    Particle & $E^*$ \\
    \hline
    $e^-$ & $23 \, a^{2/3}(\si{\micro\metre})$ \\
    H     & $133 \, a(\si{\micro\metre})$ \\
    He    & $222 \, a(\si{\micro\metre})$ \\
    C     & $665 \, a(\si{\micro\metre})$ \\
    N     & $665 \, a(\si{\micro\metre})$ \\
    O     & $665 \, a(\si{\micro\metre})$ \\
    \hline
  \end{tabular}
  \caption[Grain critical energy]{Grain critical energy, $E^*$, for a dust grain of $a$ in \si{\micro\metre} for electrons, $e^-$, as well as the elements considered for grain cooling. The values for carbon, oxygen and nitrogen are identical \citep{dwek_infrared_1981}.}
  \label{tab:p2-criticalenergy}
\end{table}

\section{System parameters}
\label{sec:paper2-wr140}

We have previously simulated WCd systems in the form of a parameter space exploration, in order to discern which wind and orbital parameters are influential on the dust growth rates \citep{eatsonExplorationDustGrain2022}.
It was determined that the primary factors of dust growth in a WCd system were the mass loss rates, $\mdot$, and wind terminal velocities, $v^\infty$, for each star, as well as the orbital separation, $d\rms{sep}$.
In particular, it was found that imbalances between the wind velocity produced Kelvin-Helmholtz (KH) instabilities due to a shear in the winds.
Slower winds were found to be more radiative in the post-shock WCR flow, cooling to temperatures suitable for dust formation.
This was found to influence the dust growth rate by as much as six orders of magnitude through a factor of four variation of the WR wind terminal velocity.
We also found that increasing $d\rms{sep}$ significantly reduced the dust production rate, due to less radiative cooling and less shock compression as the out-flowing winds became less dense with distance.
In the case of WCd systems with eccentric orbits, the separation distance can vary significantly.
In the case of WR140, $\dsep$ varies by a factor of 19 from apastron to periastron.
This is hypothesised to be the primary cause of dust production variability within episodic systems.
The pre-shock velocity can also vary somewhat due to radiative inhibition and orbital motion.

In order to understand the structure and dynamics of the CWB system we must define some important parameters, such as the wind momentum ratio, $\eta$, which is defined as:

\begin{equation}
  \eta = \frac{\mdot\sob v^\infty\sob}{\mdot\swr v^\infty\swr} .
\end{equation}

\noindent
As $\eta$ decreases we find that the wind becomes more imbalanced.
In the case of WR+OB CWB systems we find that the WR stars wind typically dominates the WCR.
Assuming that there is no radiative inhibition \citep{stevens_stagnation-point_1994} or radiative braking \citep{gayley_sudden_1997}, we can approximate the WCR to a conical region with an opening angle about the contact discontinuity:

\begin{equation}
  \theta\rms{c} \simeq 2.1 \left( 1 - \frac{\eta^{2/5}}{4}\right) \eta^{-1/3} ~~~ \text{for} ~ 10^{-4} \leq \eta \leq 1 ,
\end{equation}

\noindent
to a relatively high degree of accuracy \citep{eichler_particle_1993}.
Another important value for determining the nature of the WCR is the cooling parameter, $\chi$, which is the ratio of the time taken for the shocked wind to completely cool to the time taken for the wind to escape the shocked region:

\begin{equation}
  \label{eq:p2-chi}
  \chi = \frac{t_\text{cool}}{t_\text{esc}} \approx \frac{v_{8}^4 d_{12}}{\dot{\text M}_{-7}} , 
\end{equation}

\noindent
where $v_{8}$ is the wind terminal velocity in units of $10^8 \, \si{\centi\metre\per\second}$, $d_{12}$ is the separation distance in units of $10^{12} \, \si{\centi\metre}$ and $\mdot_{-7}$ is the wind mass loss rate in units of $10^{-7} \, \si{\solarmass\per\year}$ \citep{stevens_colliding_1992}.
As $\chi$ decreases, the structure of the WCR becomes more influenced by cooling.
If $\chi < 1$, the WCR is completely dominated by radiative cooling, while if $\chi \gg 1$, the WCR behaves adiabatically.
If the WCR is highly radiative the total compression can be significantly greater than the adiabatic limit of 4 for a $\gamma = 5/3$ gas, which facilitates dust production.
Finally, we define a maximum dust production rate of the system, $\mdot\rms{d,max}$, assuming a 100\% conversion rate of WR wind in the WCR into dust.
The fraction of the WR wind that is passed through the WCR is given by the equation:

\begin{equation}
  f\rms{WR} = \frac{1 - \cos\left(\theta\rms{WR}\right)}{2},
\end{equation}

\noindent
where $\theta\rms{WR}$ is the opening angle of the WR shock front ($\theta\rms{WR} \approx 2 \tan^{-1}(\eta^{1/3}) + \pi/9$, see \cite{pittardCollidingStellarWinds2018}).
$\mdot\rms{d,max}$ is then calculated with the formula

\begin{equation}
  \label{eq:p2-maxdustrate}
  \mdot\rms{d,max} = \mdot\rms{WR} X\rms{C,WR} f\rms{WR},
\end{equation}

\noindent
where $X\rms{C,WR}$ is the carbon mass fraction in the WR star.

\subsection{WR140 parameters}

WR140 was simulated in this paper as it is an archetypical example of an episodic WCd system.
The system has an extremely eccentric orbit, which significantly affects the cooling parameter as the orbit progresses, and is also observed in detail.
Both stellar winds are expected to be accelerated to close to their terminal wind velocities before they collide \citep{lamersIntroductionStellarWinds1999}.

Recent improved estimations of the orbital parameters of WR140 by \cite{thomasOrbitStellarMasses2021} were used to calculate the orbit for these simulations, while the mass loss rate, and the wind terminal velocity were taken from \cite{williamsMultifrequencyVariationsWolfrayet1990}
(see Table \ref{tab:wr140systemparameters}).
A typical wind composition for WC stars was assumed for the Wolf-Rayet star, while a solar abundance was assumed for the OB star (Table \ref{tab:p2-abundances}).

\begin{table}
  \centering
  \begin{tabular}{lll}
    \hline
    Parameter & Value & Citation \\
    \hline
    $\text{M}_\text{WR}$ & \SI{10.31}{\solarmass} & \cite{thomasOrbitStellarMasses2021} \\
    $\text{M}_\text{OB}$ & \SI{29.27}{\solarmass} & \cite{thomasOrbitStellarMasses2021} \\
    $P$ & \SI{7.926}{\year} & \cite{thomasOrbitStellarMasses2021} \\
    $e$ & 0.8993 & \cite{thomasOrbitStellarMasses2021} \\
    $\dot{\text{M}}_\text{WR}$ & \SI{5.6e-5}{\solarmass\per\year} & \cite{williamsMultifrequencyVariationsWolfrayet1990} \\
    $\dot{\text{M}}_\text{OB}$ & \SI{1.6e-6}{\solarmass\per\year} & \cite{williamsMultifrequencyVariationsWolfrayet1990} \\
    $v^\infty_\text{WR}$ & \SI{2.86e3}{\kilo\metre\per\second} & \cite{williamsMultifrequencyVariationsWolfrayet1990} \\
    $v^\infty_\text{OB}$ & \SI{3.20e3}{\kilo\metre\per\second} & \cite{williamsMultifrequencyVariationsWolfrayet1990} \\
    $\eta$ & 0.031 & Calculated \\
    $\chi_\text{WR,min}$ & 2.12 & Calculated \\
    \hline
  \end{tabular}
  \caption[WR140 system parameters]{The system parameters for the WR140 system as used in this paper. References for each parameter are provided.}
  \label{tab:wr140systemparameters}
\end{table}

\begin{table}
  \centering
  \begin{tabular}{lll}
  \hline
  Element & Solar & WC \\ \hline
  $X\rms H   $ & $0.705$ & $0.000$ \\
  $X\rms{He} $ & $0.275$ & $0.546$ \\
  $X\rms C   $ & $0.003$ & $0.400$ \\
  $X\rms N   $ & $0.001$ & $0.000$ \\
  $X\rms O   $ & $0.010$ & $0.050$ \\
  \hline
  \end{tabular}
  \caption[Abundances by mass used for OB and WR stars]{Abundances used for the OB and WR stars being simulated. Other elements are assumed to be trace when calculating dust emission \citep{williamsSpectraWC9Stars2015,sanderGalacticWCStars2012,andersAbundancesElementsMeteoritic1989}}.
  \label{tab:p2-abundances}
\end{table}

\subsection{Simulation parameters}
\label{sec:p2-sysparameters}

\begin{figure}
  \centering
  \includegraphics[width=\linewidth]{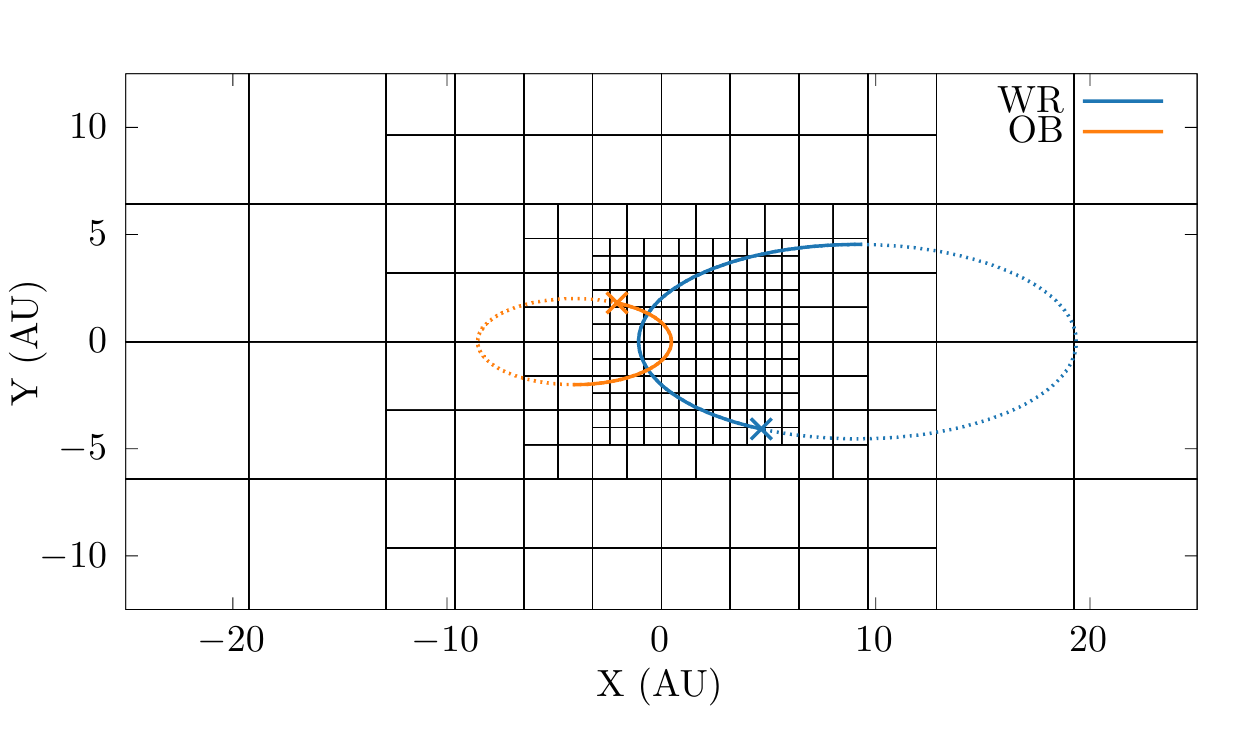}
  \caption[Numerical grid of the WR140 system simulation at $z=0$]{Numerical grid of the WR140 system simulation at $z=0$ near the maximally refined region. Static mesh refinement was used to increase the resolution around the orbital path from $0.95 \leq \Phi \leq 1.10$. The orbital path of both stars are overlaid onto this numerical grid. The stars are typically in the 3\ts{rd} or 4\ts{th} level.}
  \label{fig:p2-orbitalpath}
\end{figure}

A domain of $128 \times 128 \times 16 \, \si{\au}$ was used for this simulation, with a coarse (0\ts{th} level) grid resolution of $400\times 400 \times 50$ cells in the XYZ domain.
The simulation has 4 refinement levels, corresponding to an effective resolution of $6400 \times 6400 \times 800$ cells and a cell size of $0.02^3 \, \si{\au\cubed}$.
At periastron passage this results in $\sim 80$ cells between the stars, which was found to be enough to adequately resolve the WCR. 
This simulation has an XYZ aspect ratio of 8:8:1 in order to reduce processing time, as the bulk of dust formation was expected to occur a short distance from the WCR.
A section of the systems orbit, corresponding to an orbital phase of $0.95 \leq \Phi \leq 1.10$ was simulated.
This represents \num{1.2} year section of the systems orbit, and is the orbital period where much of the dust forms - prior to and shortly after periastron passage \citep{crowther_dust_2003}.
Fig. \ref{fig:p2-orbitalpath} shows the orbit overlaid onto the statically refined numerical grid.
The area of maximum refinement is around the orbital paths of the stars from $0.94 \leq \Phi \leq 1.11$, in order to ensure that the stars and the apex of the WCR are maximally refined.
Prior to periastron passage the \texttt{rk3} integrator was used for its speed, but increasing numerical instability as the stars grew closer resulted in this proving untenable, and it was switched to \texttt{ssprk5\_4} for the remainder of the simulation.

During periastron passage the average time-step was found to reduce by an order of magnitude, resulting in a corresponding increase to simulation time. %
At the most numerically complex portion of the simulation, a Courant number of $C = 0.04$ had to be used instead of the initial value of $C = 0.15$, in order to preserve numerical stability.

\subsection{Data collection}
Simulation data was exported as HDF5 files at regular time intervals.
3D meshes were collected every increment of $\delta \Phi = \num{1.5e-3}$, while 2D slices in the XY plane were collected every increment of $\delta \Phi = \num{1.5e-4}$.
These HDF5 files contain the primitive variables of the simulation: gas density, $\rho$, gas pressure, $P$, and wind velocity components, $v_x$, $v_y$ and $v_z$.
The scalars governing the dust properties were also stored for each cell, in particular the dust-to-gas mass ratio, $z$.
The volume-weighted totals of all parameters of interest were also collected, such as the average values for the dust production rate within the WCR, $\dot{\text{M}}\rms{d}$.

\section{Results}
\label{sec:p2-results}

\begin{table}
  \centering
  \begin{tabular}{lll}
  \hline
  Parameter & Mean & Maximum \\ \hline
  $\dot{\text{M}}\rms{d}$ (\si{\solarmass\per\year}) & \num{7.68e-08} & \num{1.24e-06} \\
  $\bar{a}$ (\si{\micro\metre}) & \num{1.32e-02} & \num{1.44e-02} \\
  $\bar{z}$ & \num{3.98e-04} & \num{3.32e-03} \\ \hline
  \end{tabular}
  \caption[Advected scalar yields from WR140 simulation]{Advected scalar yields from the WR140 simulation.}
  \label{tab:paper-2-dust-rates}
\end{table}

\begin{figure}
  \centering
  \includegraphics[width=\linewidth]{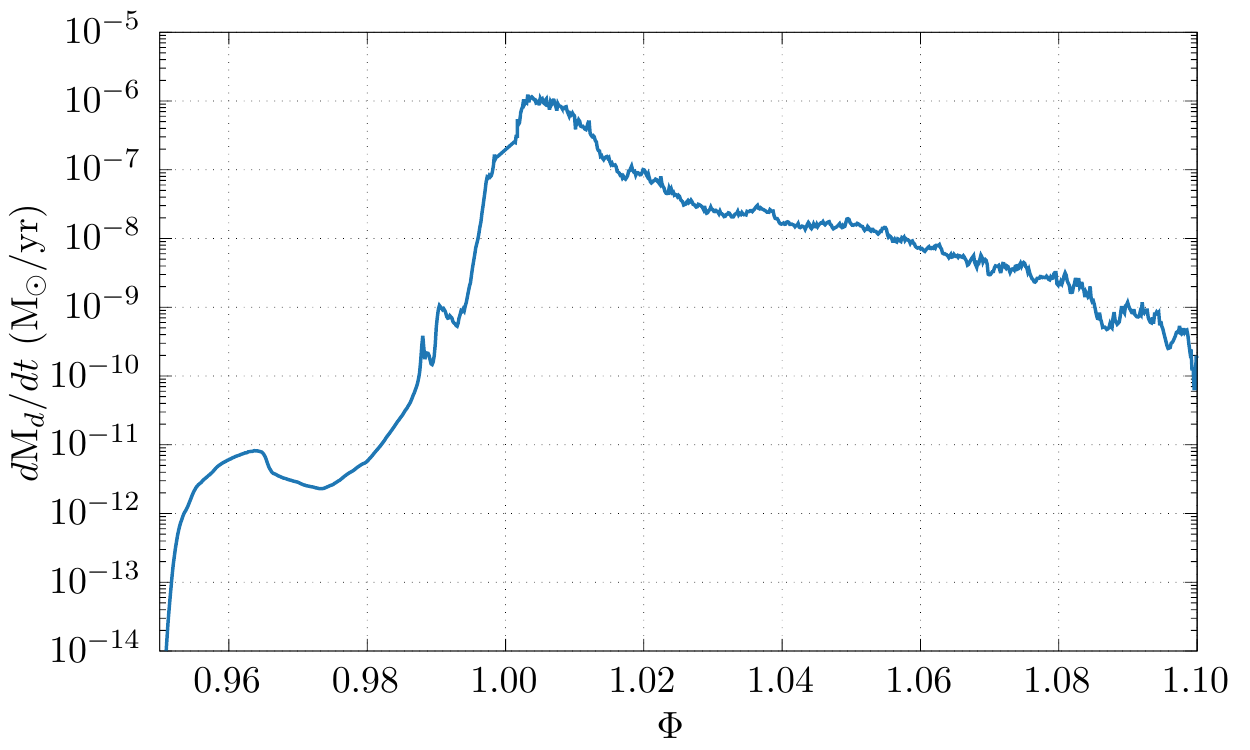}
  \caption[A graph of the dust production rate in the WCR over the orbital phase $0.95 \leq \Phi \leq 1.10$]{A graph of the dust production rate in the WCR over the orbital phase $0.95 \leq \Phi \leq 1.10$. The dust production rate sharply increases as the stars pass their closest approach. Afterwards, the dust production rate begins to falter and slow, due to weaker wind collision effects via the separation distance and radial velocity.}
  \label{fig:wr140-dustproduction}
\end{figure}

\begin{figure}
  \centering
  \includegraphics[width=\linewidth]{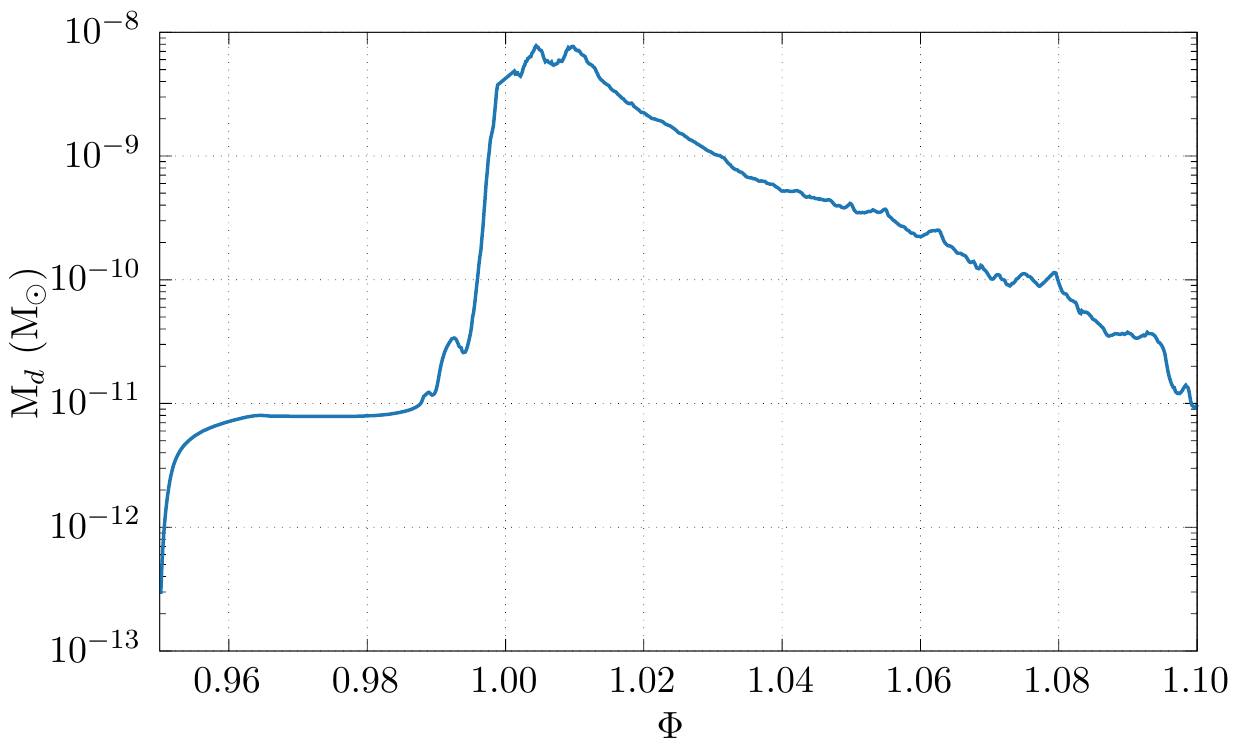}
  \caption[A graph of the overall dust mass in the simulation of WR140 over the orbital phase $0.95 \leq \Phi \leq 1.10$]{A graph of the overall dust mass in the simulation of WR140 over the orbital phase $0.95 \leq \Phi \leq 1.10$. The amount of dust quickly reduces after periastron due to a decreased dust growth rate (Fig. \ref{fig:wr140-dustproduction}), as well as dust advecting off the numerical grid. While we observe a double peak in the dust production rate, this is due to a high level of grain ablation occurring as well as grain growth.}
  \label{fig:wr140-dustmass}
\end{figure}

\begin{figure}
  \centering
  \includegraphics[width=\linewidth]{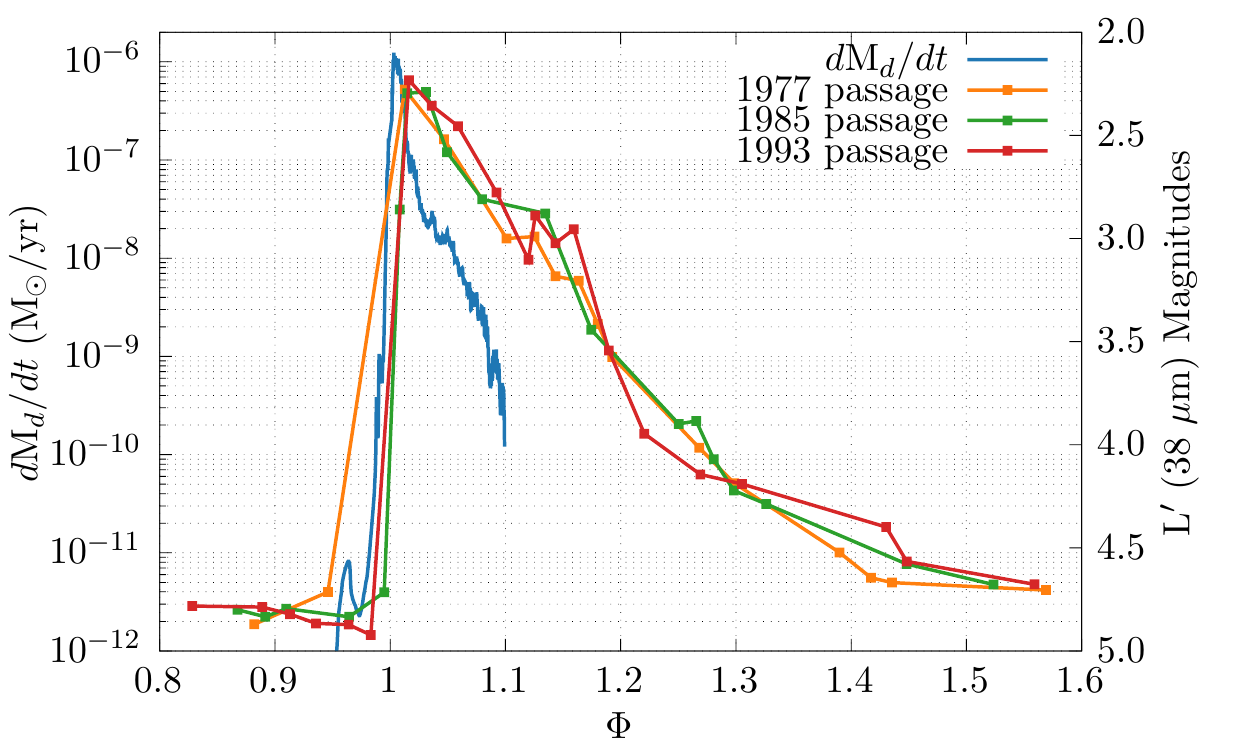}
  \caption[Comparison of simulation dust production rate and $L^\prime$ photometry]{A comparison of the dust production rate in our simulation of WR140 and $L^\prime$ photometry of the WR140 system at three periastron passages in 1977, 1985 and 1993 \citep{crowther_dust_2003}. Radiance from dust emission peaks shortly after $\Phi = 1.00$ in all cases, and reduces to regular levels by $\Phi \sim 1.50$. The dust production rate peaks at the same time and reduces in a similar manner. Whilst dust production is expected to reach pre-periastron levels sooner than $\Phi \sim 1.50$, this is not a direct comparison as dust can radiate long after leaving the dust production region. Conversion from observation date to orbital phase is performed with orbital parameters defined in \cite{fahedSpectroscopyArchetypeCollidingwind2011}.}
  \label{fig:wr140-lprime}
\end{figure}

\begin{figure}
  \centering
  \includegraphics[width=\linewidth]{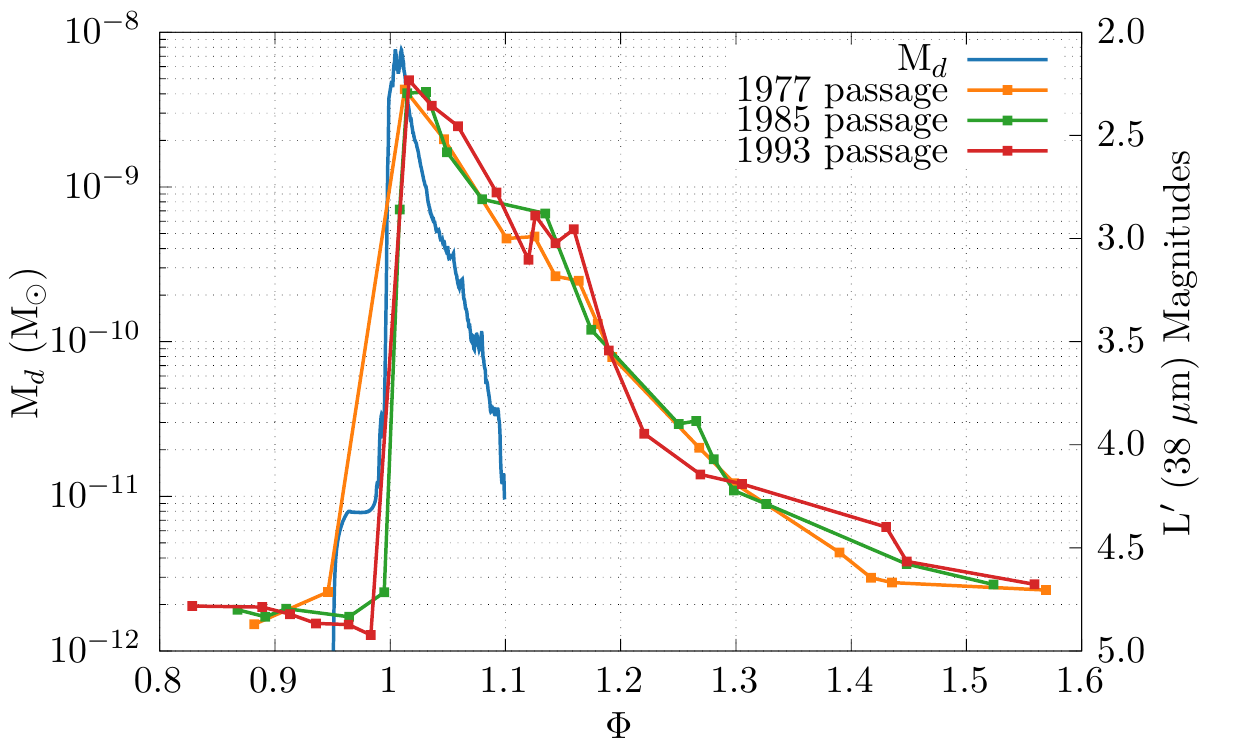}
  \caption[Comparison of simulation dust mass and $L^\prime$ photometry]{A comparison of the simulation dust mass and $L^\prime$ photometry of the WR140 system at three periastron passages in 1977, 1985 and 1993 \citep{crowther_dust_2003}. Radiance from dust emission peaks shortly after $\Phi = 1.00$ in all cases, and reduces to regular levels by $\Phi \sim 1.50$. Dust mass in the simulation reaches a maxima at the same time as dust luminosity peaks, with dust mass steadily reducing in a similar manner. The dust mass reduces to pre-periastron levels faster than the dust production rate in Fig. \ref{fig:wr140-lprime} as the dust quickly advects off of the numerical grid.}
  \label{fig:wr140-lprime-mass}
\end{figure}

We observe that as the system approaches periastron passage, the WCR temperature decreases and the WCR on the whole becomes radiatively dominated.
Instability appears on the leading shock prior to and during periastron passage.
Similarly to what was observed in our previous paper, the leading shock typically has more instability, as it sweeps through the WR stellar wind.
After the periastron passage, both the leading and trailing shock display instabilities.
We observe that these instabilities persist for some time after periastron passage, even when the value of $\chi$ suggests that the wind interaction region should be behaving adiabatically.
This is curious, and is further discussed in more detail in Section \ref{sec:p2-instab}.

Dust growth was found to be consistent with previous uses of this particular dust model, and well below the theoretical maximum dust formation rate, $\mdot\rms{d,max} \approx \SI{4.8e-6}{\solarmass\per\year}$.
After an initial adjustment phase lasting until $\Phi \approx 0.96$, the dust production rate rapidly increased as the stars approached periastron passage, peaking at $\Phi \sim 1.01$ (Fig. \ref{fig:wr140-dustproduction}).
This maximum dust production rate of \SI{1.24e-6}{\solarmass\per\year} is sensible, but incredibly prodigious.
We find peak conversion efficiency of gas into dust of $\sim 26\%$ in the WCR compared to the theoretical maximum dust production rate described in Eq. \ref{eq:p2-maxdustrate}, as well as a total conversion efficiency of $\sim 2.2\%$ throughout the entire system (assuming a total mass loss rate of \SI{5.76e-5}{\solarmass\per\year}).
The majority of dust ($\gtrsim 90\%$) is produced within the WCR, with only a very small amount being produced outside the WCR.
Most, if not all of this dust is produced near to the apex of the WCR shock, with dust growth being less prevalent further away from this apex.
The dust produced outside of the WCR is likely an artefact of the initial injection of dust into the simulation.
After reaching this maximum value, the dust production rate steadily decreases as the stars recede from each other.
Assuming this decrease in dust production rate is steady, dust production reaches a pre-periastron level of $\sim 10^{-11} \, \si{\solarmass\per\year}$ by $\Phi \lesssim 1.15$, giving us an active dust production period of $\lesssim 1.2 \, \si{\year}$.
This is consistent with $L^\prime$ photometry of WR140 \citep{crowther_dust_2003} -- the photometry magnitude sharply increases over the first year -- and suggests our model can qualitatively assess dust production rates of episodic WCd systems.
This is reflected in the overall dust mass of the simulation (Fig. \ref{fig:wr140-dustmass}), as well as in infrared observations of WR140, where the infrared emission due to dust rapidly reaches a maximum value after periastron passage, and slowly relaxes to a minimum value.
In the case of Fig. \ref{fig:wr140-dustmass} we observe a double peak structure between $\Phi = 1.00$ and $\Phi = 1.02$.
At the peak rate of dust formation over this period, we would observe a dust mass in the simulation approaching \SI{6e-8}{\solarmass} if dust destruction were not included in the simulation.
As such, we can infer that the peak structure is due to a pronounced increase in the dust destruction rate as the post-shock density and temperature are at their highest during the systems periastron passage.

Fig. \ref{fig:wr140-lprime} compares the dust production rate in the simulation and available $L^\prime$ photometry of WR140, we observe that the dust production rate of the simulation drops in a similar manner to the drop in dust emission in the system after periastron passage.
Whilst the rate of decay of dust production rate is much faster than the rate of decay of $L^\prime$ luminance, this is not a direct comparison -- dust is still capable of radiating after leaving the region of the WCR where dust growth occurs most intensely.
Fig. \ref{fig:wr140-lprime-mass} shows a similar result, though the dust mass reduces significantly faster as the dust rapidly traverses through the simulation and off and out of the numerical domain.
This asymmetry in the time-dependent change in infrared luminosity implies the existence of several factors for suppression and encouragement of dust formation than just the change in orbital separation distance.
These results are also in agreement with more contemporary research by \cite{pollockCompetitiveXRayOptical2021} which notes a similar sharp increase in K-band flux and x-ray column density a short time after periastron passage.
This increase in flux and column density corresponds to the rapid production of dust grains, which subsequently cool as they recede from the shock region.
Furthermore, \cite{pollockCompetitiveXRayOptical2021} notes that x-ray cooling becomes less influential, and instead transitions to an optical cooling regime shortly after periastron passage, indicating a rapid rate of cooling in the immediate post-shock environment.

The evolution of dust in this system would result in the formation of an expanding cloud of dust every time the system passes periastron, with no contiguous spiral pattern forming, due to the lengthy ``dormant'' period occurring shortly after periastron passage.
This is consistent with observations of WR140, where these disconnected clouds are observed \citep{williams_orbitally_2009}.
We find an average dust production rate of $\mdot\rms{d} = \SI{7.68e-8}{\solarmass\per\year}$ over the entire simulation, and a change in the dust production rate by approximately five orders of magnitude over the course of the simulation.
This fits our understanding of an episodic dust forming WCd system, with an extremely clear ``active'' period followed by a slow tapering off of dust production as the system approaches the ``dormant'' period.
Our value for the dust-to-gas mass ratio within the system appears to be sensible, while our average dust production rate is significantly higher.
This is due to the limited temporal sample of the simulation.
We would find a significantly lower average dust production rate over the course of a full orbit due to more sampling of the system over the ``dormant'' period.
We can compare our results to the estimated dust yields from \cite{lauRevisitingImpactDust2020}, which found an average dust production rate of $\mdot\rms{d} = \SI{8.11e-10}{\solarmass\per\year}$ throughout an entire orbit of WR140.
Extrapolating our results with a dust production period of \SI{1.2}{\year} we find an average dust production rate of $\sim 10^{-8} \, \si{\solarmass\per\year}$ over the total orbit.
This is within an order of magnitude of the estimated dust production rate from \cite{lauRevisitingImpactDust2020}, though indicates that the model requires additional work in order to perform quantitative analysis of these systems -- which is in line with our conclusions in our previous paper.

\begin{figure*}
  \centering
  \includegraphics[width=0.95\linewidth]{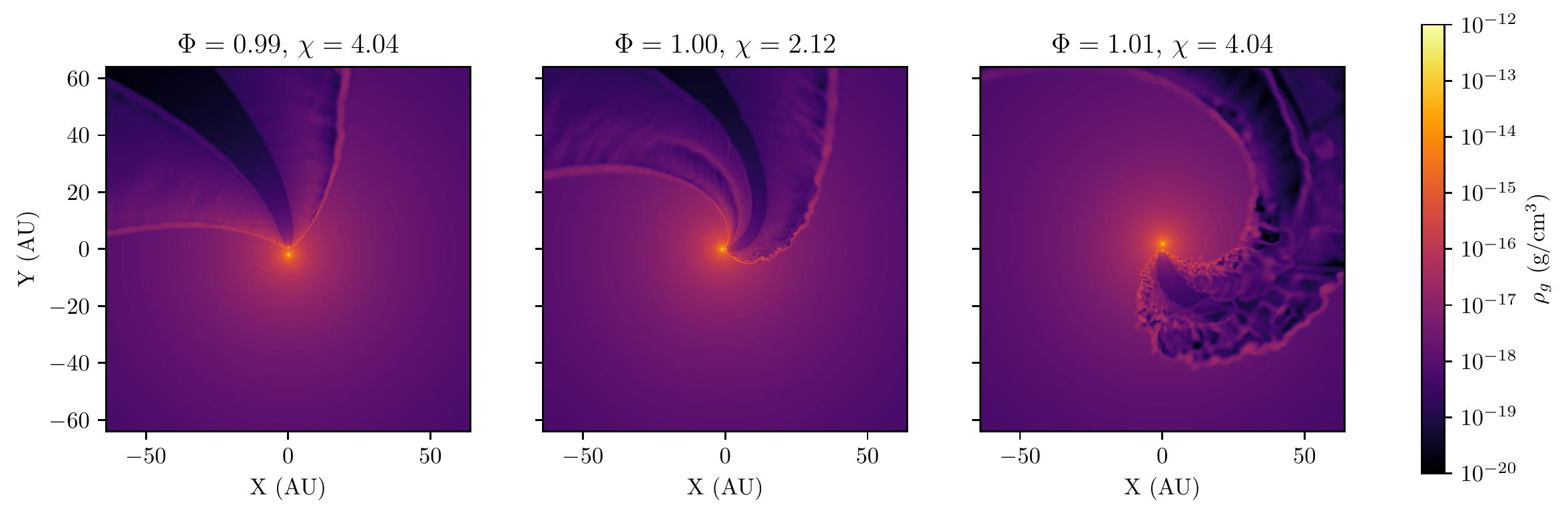}
  \caption[Gas density in a simulation of the WR140 system]{Gas density in a simulation of the WR140 system shortly before, during, and after periastron. The value of $\chi\rms{WR}$ is calculated using Eq. \ref{eq:p2-chi} and the WR wind parameter is noted in each panel. The simulation becomes dominated by instabilities after periastron. These instabilities persist despite the system behaving adiabatically at $\Phi = 0.99$, when the  orbital separation distance is identical. This suggests that the radiative behaviour of the post-shock WCR is due to other factors, in addition to $d\rms{sep}$.}
  \label{fig:p2-fullpage-rho}
\end{figure*}

\begin{figure*}
  \centering
  \includegraphics[width=0.95\linewidth]{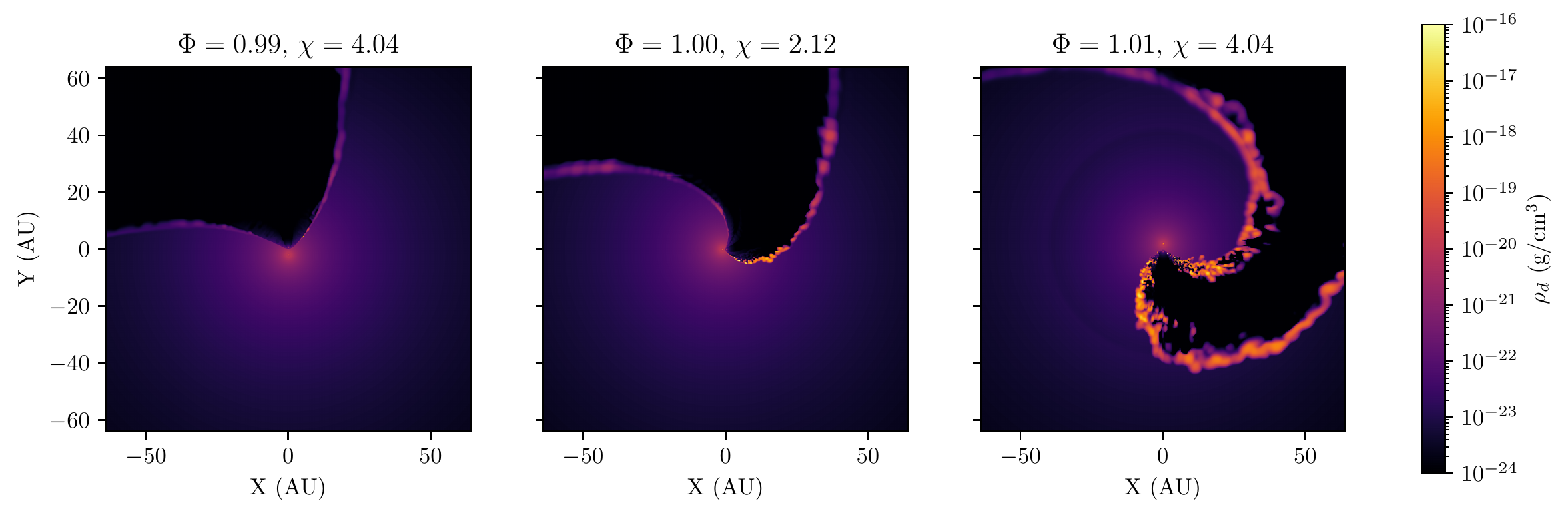}
  \caption[Dust density in a simulation of the WR140 system]{Dust density in a simulation of the WR140 system shortly before, during, and after periastron. Dust growth occurs as a direct result of the formation of cold, dense gas in the post-shock WCR.}
  \label{fig:p2-fullpage-rhod}
\end{figure*}

\begin{figure*}
  \centering
  \includegraphics[width=0.95\linewidth]{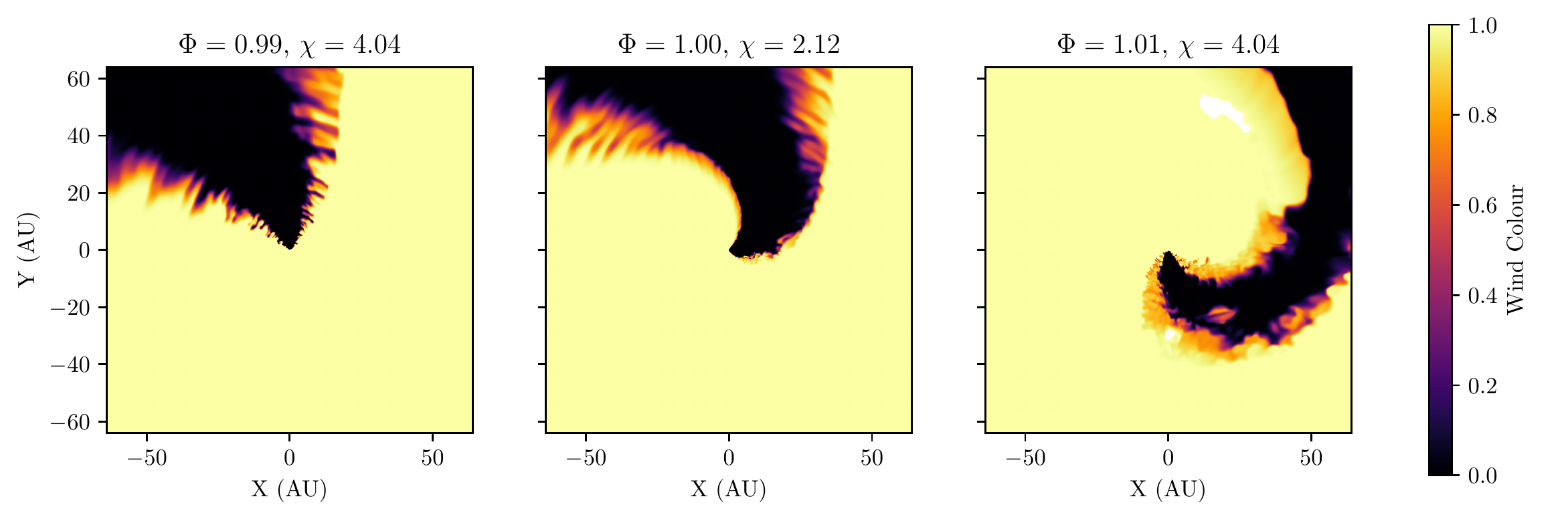}
  \caption[Wind ``colour'' in a simulation of the WR140 system]{Wind ``colour'' in a simulation of the WR140 system shortly before, during, and after periastron. A colour of 1 represents a pure WR wind and a colour of 0 represents a pure OB wind. We find that the wind undergoes more mixing during and after periastron.}
  \label{fig:p2-fullpage-r0}
\end{figure*}

\subsection{Instabilities}
\label{sec:p2-instab}

As can be seen in Fig. \ref{fig:p2-fullpage-rho}, after periastron passage the post-shock WCR region transitions from a smooth adiabatic wind to a highly radiative wind dominated by radiative, thin-shell instabilities.
As the WCR becomes increasingly radiative, dust growth drastically increases, with the bulk of dust production occurring within the high density regions produced by these instabilities.
However these clumpy pockets of gas do not exhibit significant dust growth beyond $\sim \SI{20}{\au}$ from the simulation barycentre, with concentrations of dust remaining approximately constant (Fig. \ref{fig:p2-fullpage-rhod}).
KH instabilities can also be present, and have been observed to result in significantly increased dust formation \citep{eatsonExplorationDustGrain2022}, the ratio of wind terminal velocities for WR140 is comparatively low.
Therefore, radiative instabilities are the primary motivator driving dust formation in WR140.
By the end of the simulation at $\Phi = 1.10$, the WCR is still somewhat dominated by instabilities, with an elevated dust production rate even though the cooling parameter has increased significantly to $\chi = 19.7$, which would normally imply adiabatic behaviour.
Whilst the dust production rate has reduced significantly, there is still a significantly greater growth rate than at the start of the simulation (after advection).
Clearly the transition from radiative back to adiabatic behaviour has a degree of latency, with instabilities still driving the structure of the WCR long after an adiabatic flow should have re-established.
Such behaviour was first reported by \cite{pittard_3d_2009} and leads to hysteresis in the x-ray emission \citep{pittard_3d_2010}.
This is the key physics responsible for the asymmetric dust growth rate about periastron.
It seems that once instabilities form in the WCR, they are somewhat difficult to stop - and cause oblique shocks, with lower post-shock temperature and ultimately stronger cooling than expected.
The amount of wind being mixed in the WCR also significantly increases after periastron passage, which would be conducive to the formation of complex organic molecules on the surface of the dust grains (Fig. \ref{fig:p2-fullpage-r0}).
Whilst this is beyond the scope of the current work, evolution of dust grains from WCd systems on longer time and length scales would be an enlightening avenue of research.

\subsection{Influence of varying wind velocity on dust production}

\begin{figure}
  \centering
  \includegraphics[width=\linewidth]{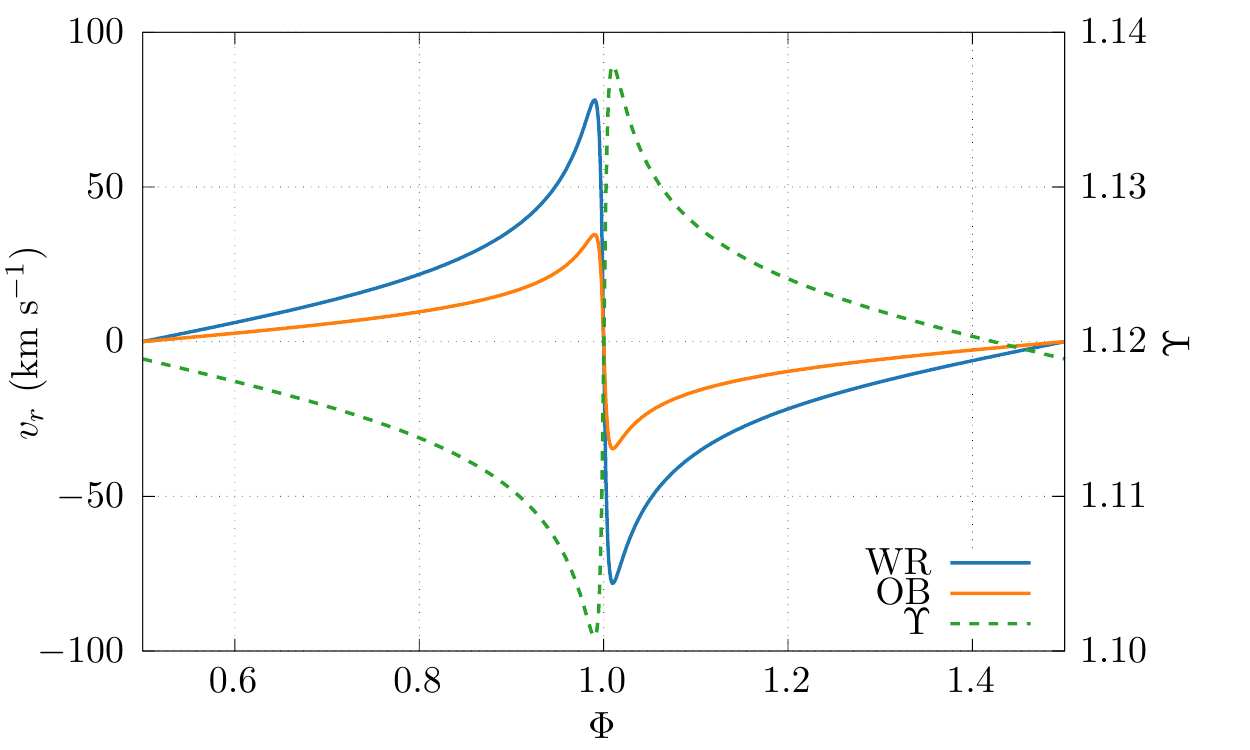}
  \caption[Radial velocity as a function of the orbital phase for the WR and OB stars in the WR140 system]{Radial velocity as a function of the orbital phase for the WR and OB stars in the WR140 system relative to the barycentre. As periastron passage occurs, the sudden inversion from approaching to receding can alter pre-shock the wind velocity of the WR star by as much as \SI{160}{\kilo\metre\per\second}. Whilst this is $\sim 6\%$ of the WR wind velocity, it can significantly increase dust production when the stars recede from each other. The velocity shear, $\Gamma = v\rms{OB}/v\rms{WR}$, also sharply increases during periastron passage, peaking at the point of maximum dust production.}
  \label{fig:p2-shear}
\end{figure}

As we have previously discussed, varying the wind terminal velocity for both stars in a simulation can result in significant changes in the dust production rate.
This is theorised to be due to the increased cooling in the WCR caused by slower winds, as well as through KH instabilities driven by a wind velocity shear (if the wind terminal velocities are significantly different, see \cite{stevens_colliding_1992}).
Previous work on this subject considered systems with circular orbits \citep{eatsonExplorationDustGrain2022}, where the pre-shock velocities are constant over the orbit of the system.
However, in the case of a system with an eccentric orbit (such as WR140), the outflow velocity for each wind, as well as their ratio, can markedly vary over the systems orbit.
As the stars approach periastron, the radial velocity, $v\rms{r}$, for each star rapidly changes from a maximum value to a minimum, as the stars approach and then swing past one another (Fig. \ref{fig:p2-shear}).
This causes a rapid change in the pre-shock velocity for both winds.
This then influences the amount of radiative cooling in the post-shock wind, suppressing radiative cooling pre-periastron and inciting it post-periastron; leading to changes in the dust production rate. 
While the change in wind velocity is relatively small (with the wind velocity varying by as much as 6\% over the course of an orbit) it still impacts the cooling of the system, and can vary $\chi$ by as much as a factor of 1.26 in the case of WR140.

The rate of dust production is also affected by the presence of a large wind velocity ratio, $\Upsilon$, where:

\begin{equation}
  \Upsilon = v\rms{OB} / v\rms{WR} .
\end{equation}

\noindent
As the mass of each star is different, the change in radial velocity differs for each star, causing a variability in the velocity ratio and therefore velocity shear.
Previous research with dust models suggests that a strong velocity shear drives an increased dust production rate.
We find that the maximum velocity shear occurs at $\Phi = 1.01$ (Fig. \ref{fig:p2-shear}), around the same time where the dust growth rate is at a maximum; this is consistent with our previous work.
This change in velocity shear may therefore be another factor behind the increased dust growth of WR140 post-periastron.

We also note that in reality there could be significant change in the pre-shock velocity of the O wind due to the WCR encroaching into the acceleration region during periastron \citep{sugawaraSuzakuMonitoringWolf2015}.

\section{Summary}
\label{sec:p2-conclusion}

Despite only simulating a limited section of the orbit of WR140, we have made a number of insights into the behaviour of the system.
We find a significant degree of change in the dust production rate as a direct consequence of the changing orbital separation of the system.
This is related to the change in the behaviour of the post-shock WCR wind, which goes from a smooth adiabatic wind to a clumpy, high density wind dominated by instabilities ideal for dust growth.
It is particularly interesting to note that the system does not revert to behaving adiabatically as quickly as expected.
This suggests that the post-shock WCR condition of the system is dependent on additional factors, instead of being solely due to $d\rms{sep}$.
One of the factors on this delayed return to the adiabatic state is potentially due to the orbital motion of the stars themselves.
As the stars approach each other at periastron, the radial velocity of the stars adds velocity to the wind beyond the outflow velocity, resulting in higher wind collision velocities, which encourages adiabatic behaviour in the post-shock flow.
The inverse is true as the stars recede from one another: the effective wind velocity for both stars is reduced, which encourages cooling and the formation of instabilities in the WCR.
Furthermore, an increased wind velocity ratio occurs after periastron due to the orbital dynamics which can drive KH instabilities.

\section*{Acknowledgements}

This work was undertaken on ARC4, part of the High Performance Computing facilities at the University of Leeds, UK.
We would also like to thank P. A. Crowther for his work on the Galactic Wolf-Rayet Catalogue (\url{http://pacrowther.staff.shef.ac.uk/WRcat}).
Finally, we would like to thank the academic referee for their extremely helpful review.

\section*{Data Availability}
The data underlying this article are available in the Research Data Leeds Repository, at \url{https://doi.org/XXXX}.

\bibliographystyle{mnras}
\bibliography{references.bib} %

\bsp	%
\label{lastpage}
\end{document}